\documentclass[twocolumn,twoside,slac_two]{revtex4}
\usepackage{graphicx}
\usepackage{fancyhdr}
\usepackage{hyperref}
\hypersetup{
    colorlinks=true,
    urlcolor=magenta,
}
\begin{document}

\title{Computation of ion production rate and short, mid and long term ionization effect by cosmic rays during Bastille day event}

%

\author{A. Mishev}
\affiliation{Space Climate Research Unit, University of Oulu, Oulu, Finland}
\author{P.I.Y. Velinov}
\affiliation{Institute for Space Research and Technology, Bulgarian Academy of Sciences, Sofia, Bulgaria}

\begin{abstract}
The galactic cosmic rays are the main source of ionization in the Earth stratosphere and troposphere. The induced by primary cosmic ray particles ionization is important in various processes related to atmospheric physics and chemistry, specifically the minor constituents. 
The ion production in the atmosphere is enhanced compared to the average following major solar energetic particles events, specifically over the polar caps. During the solar cycle 23 we observed several strong ground level enhancements, one of the strongest among them been  the Bastille day event on 14 July 2000. In the work presented here we apply a full Monte Carlo 3-D model in order to compute the cosmic ray induced ionization. The model is based on atmospheric shower simulation with the PLANETOCOSMICS code and the ion production rate is considered as a superposition of cosmic rays with galactic and solar origin. The ion production rate is computed as a function of the altitude above the sea level and the short, mid and long term ionization effect relative to the average due to galactic cosmic rays is computed.
\end{abstract}

\maketitle

\thispagestyle{fancy}


\section{INTRODUCTION}
The possible effect of high energy cosmic ray particles on various atmospheric processes related to atmospheric chemistry
and physics is debated over the last years. Recent findings suggest an apparent influence of cosmic rays on various atmospheric processes and electric circuit, as well as on minor constituents of the atmosphere \cite{Bazilevskaya08}, \cite{Mironova2015}. Up to present, in most of the proposed and debated models, the induced by cosmic rays, both from galactic and/or solar origin  atmospheric ionization plays a key role. 

Our planet Earth is constantly bombarded by high, very high and ultra-high energy nuclei, known as cosmic rays, which are the main source of ionization
in the troposphere \cite{Usoskin09}. The contribution of this particles to the atmospheric
ionization is continuous with slight variation in time due to modulation effects in the Heliosphere. Occasionally solar energetic
particles (SEPs) enter the Earth atmosphere, penetrate deep into in the atmosphere or even reach the surface, in a such way leading to ground level enhancements (GLEs). 
As a result they cause an excess of ionization, specifically over the polar caps \cite{Usoskin11a}, \cite{Mishev2013}. 

At recent as a result of numerical methods, based on enhanced knowledge of high-energy interactions and nuclear processes several models for estimation of cosmic ray induced ionization in the Earth atmosphere have been proposed within good agreement with experimental results \cite{Desorgher05}, \cite{Usoskin06}, \cite{Velinov2009}, \cite{Wissing2009}. 

These models are based on a full Monte Carlo simulation of the atmospheric cascade. As was recently reviewed they agreed within  10--20 $\%$ \cite{Usoskin09}. These full target models allow one to compute the ion production rate, accordingly ionization effect in the atmosphere during major GLEs as superposition of the contribution of cosmic rays with galactic and solar origin \cite{Mishev13b}, \cite{Mishev15}. Here we present the results of computation of ion production rate and corresponding ionization effect relative to the average due to galactic cosmic rays during one of the most interesting and
major events of the previous 23 solar cycle, namely the GLE 59 on Bastille day of 14 July 2000.

\section{MODEL}
Here we use model similar to \cite{Usoskin06}, the full description given elsewhere \cite{Mishev07, Velinov2009}. The ion production rate is given by: 

\begin{equation}
  q(h,\lambda_{m}) =  \frac{1} E_{ion} \sum_{i} \int_{E_{cut}(R_{c})}^{\infty} \int_{\Omega} D_{i}(E) \frac{\partial E(h,E)}{\partial h} \rho(h)dE d\Omega 
        \label{simp_eqn1}
   \end{equation}

\noindent where $\partial E$ is the deposited energy in an atmospheric layer $\partial h$, $h$ is the air overburden above a given altitude in the
atmosphere expressed in $g/cm^{2}$ subsequently converted to altitude above the sea level (a.s.l.), $D_{i}(E)$ is the differential cosmic ray spectrum for a given
component $i$: protons p, Helium ($\alpha$-particles), Light nuclei L (3 $\le$ Z $\le$ 5), Medium nuclei M (6 $\le$ Z $\le$ 9), Heavy nuclei H (Z $\ge$ 10) and Very Heavy nuclei VH (Z $\ge$ 20) in the composition of primary CR nuclei (Z is the atomic number), $\rho$ is the atmospheric density in $g.cm^{-3}$, $\lambda_{m}$ is the geomagnetic latitude, $E$ is the initial energy of the incoming primary nuclei on the top of the atmosphere, $\Omega$ is the geometry factor - a solid angle and $E_{ion}$ = 35 eV is the energy necessary for creation of an ion pair in air \cite{Porter76}. The integration is over the kinetic energy above $E_{cut}(R_{c})$, which is defined by the local rigidity cut-off $R_{c}$ for a nuclei of type $i$ at a given geographic location by the expression:

\begin{equation}
 E_{cut,i}=\sqrt{ \left( \frac {Z_{i}} {A_{i}}\right)^{2} R_{c}^{2}+ E_{0}^{2}} - E_{0}
        \label{simp_eqn2}
   \end{equation}
  
\noindent where $E_{0}$ =  0.938 GeV/n is the proton's rest mass. Accordingly, for SEPs spectra in equation (1), which are considerably varying from event to event, we consider results derived on the basis of ground based measurements with neutron monitors.  In this study, the propagation and interaction of high energy protons with the atmosphere are simulated with the PLANETOCOSMICS code \citep{Desorgher05} assuming a realistic atmospheric model NRLMSISE2000 considering seasonal influence \cite{Picone02} \cite{Mishev08}, \cite{Mishev2014}. PLANETOCOSMICS provides detailed simulation of particle interaction with atmosphere in a wide range of energy and altitudes with a very good resolution and allows one to simulate realistically the interactions and, when appropriate, decay of nuclei, hadrons, muons, electrons and photons in the atmosphere. In addition to the detailed detailed information about the flux of secondary particles at a given atmospheric depth it provides the energy loss and deposition, necessary for the computations with ~Eq. (\ref{simp_eqn1}). Therefore the model allows one to estimate the ion production rate, accordingly the ionization effect  in a whole atmosphere.

\section{Ion production rate during the Bastille day GLE 59}
As mentioned above, the ion production rate during major GLEs is a superposition of the contribution of galactic cosmic rays (GCR) and GLE particles, which typically  possess an essential anisotropic part. Therefore, it is necessary to compute the rigidity cut-off at given geographic position and to apply the described above model using an appropriate model for GCR spectrum as well as to consider explicitly the anisotropy by computation of the asymptotic cones in the region of interest. 

 \begin{figure}
\centering \resizebox{\columnwidth}{!}{\includegraphics{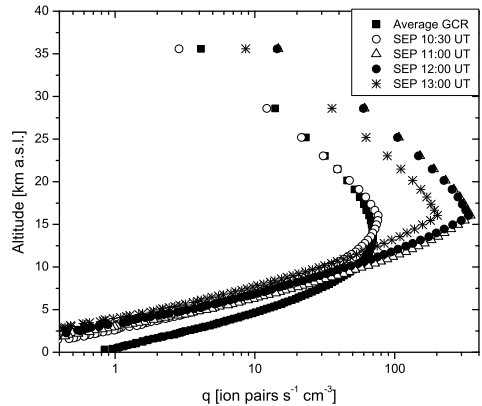}}
 \caption{Ion production rate during the Bastille day GLE on 14 July 2000 in the polar and sub-polar region with rigidity cut-off $R_{c}$ $\le$ 1 GV \label{fig1}}
 \end{figure}

 \begin{figure}
\centering \resizebox{\columnwidth}{!}{\includegraphics{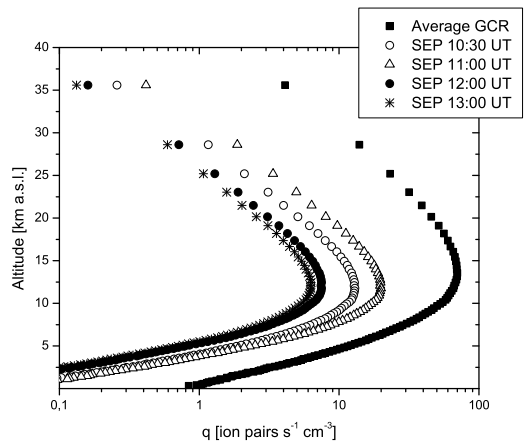}}
 \caption{Ion production rate during the Bastille day GLE on 14 July 2000 in the polar and sub-polar region with rigidity cut-off $R_{c}$ $\le$ 2 GV \label{fig1}}
 \end{figure}

The mid July 2000  was characterized by intense solar activity, resulting on three X-class solar flares (including the Bastille Day flare) and two halo coronal mass ejections (CME) \cite{Dryer2001}. The GLE 59 event was related to the Bastille day X5.8/3B solar flare and the associated full halo
CME. The event started at 10:03 UT, reached peak at 10:24 UT and ended at 10:43 UT \cite{Klein2001a}. Accordingly, the GLE onset began between 10:30 and 10:35 UT at several
stations with strongest NM increases observed at the South Pole (58.3 $\%$) and SANAE (54.4 $\%$) compared to pre-increase levels. In general the event was characterized by a large anisotropy in its initial phase \cite{Bom06, Mishev16}. 

With the reconstructed spectra used as input (see Eq.~\ref{simp_eqn1}) the ion production rate during the Bastille day event on 14 July 2000 was computed at 1GV and 2 GV rigidity cut-off \cite{Mishev15a}. The time evolution of ion production due to CR of galactic and solar origin during the GLE 59 is presented in Fig. 1 for 1 GV rigidity cut-off, accordingly in Fig.2 for 2 GV rigidity cut-off. 

The computed ion production rate is significant during the main phase of the event at the polar and sub-polar region with rigidity cut-off of about 1 GV. The ion production is significant in the low stratosphere (Fig. 1). During the initial and late phases of the event the ion production is comparable to the average due to GCR, but at altitudes of about 10 km a.s.l. and below the ion production due to GCR is greater than SEPs, because of the rapidly falling spectra of the solar particles. At high mid latitudes and mid latitudes with rigidity cut-off of about 2--3 GV, the ion production due to GCR dominates in the whole atmosphere throughout the event (Fig. 2), because the very soft SEPs spectra.

\section{Ionization effect}
The computed on a step ranging from 5 to 30 min. ion production due to CR of solar and galactic origin allows one to estimate the ionization effect during the GLE 59 on 14 July 2000 at several altitudes \cite{Mishev16a}. The expected maximal ionization effect relative to the average due to GCRs at 1 GV and 2 GV rigidity cut-offs without considering the anisotropy is shown on Fig. 3. 

 \begin{figure}
\centering \resizebox{\columnwidth}{!}{\includegraphics{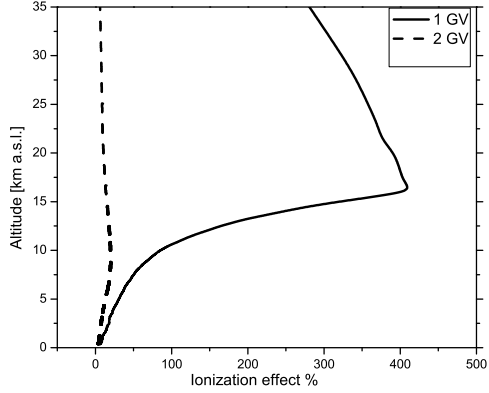}}
 \caption{Maximal ionization effect averagen over the event (5h) during the Bastille day GLE on 14 July 2000 in the region with $R_{c}$ $\le$ 1 GV and $R_{c}$ $\le$ 2 GV \label{fig1}}
 \end{figure}

Accordingly on Fig. 4 is presented the computed total 5 h ionization effect in the Earth atmosphere during the GLE 59 in the polar
and sub-polar region at altitude of about 12 km a.s.l.. The corresponding 24h ionization effect is presented on Fig. 5.

 \begin{figure}
\centering \resizebox{\columnwidth}{!}{\includegraphics{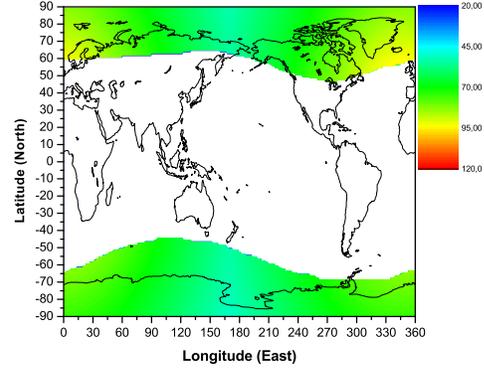}}
 \caption{Event averaged ionization effect during the Bastille day GLE on 14 July 2000 in the polar and sub-polar region with rigidity cut-off $R_{c}$ $\le$ 1 GV \label{fig1}}
 \end{figure}
 
 \begin{figure}
\centering \resizebox{\columnwidth}{!}{\includegraphics{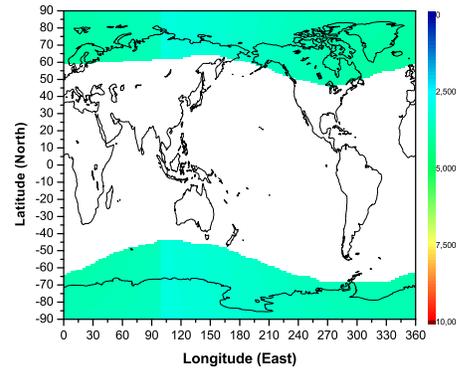}}
 \caption{24$^h$ ionization effect during the Bastille day GLE on 14 July 2000 in the polar and sub-polar region with rigidity cut-off $R_{c}$ $\le$ 1 GV \label{fig1}}
 \end{figure} 

The event averaged ionization effect at this altitude (12 km a.s.l.) is maximal (nearly 80 $\%$) in the regions 0$^{\circ}$ -- 30$^{\circ}$ E and 300$^{\circ}$◦ -- 360$^{\circ}$ E, particularly in the Northern
hemisphere. Accordingly the minimal ionization effect at altitude of 12 km a.s.l, is about 50$\%$, and it is
observed in anti-sunward direction, namely in the region 120--225$^{\circ}$ E. The computed total 24 h ionization effect as shown in Fig. 5 is not significant and it is in the order of 5$\%$. In the low troposphere is drops to about a 1$\%$.
 
\section{Conclusion}
The relative to the average due to GCR ionization effect in the polar and sub-polar regions of the
Earth (with geomagnetic cut-off rigidities between 0 GV – in the cusp, and
1 GV) is significant during the whole event, specifically in the region of the Pfotzer
maximum, thus in the lower stratosphere, the tropopause and in the upper
troposphere. In the region of high-middle latitudes, as well as in low- middle latitudes the ionization effect is weak during the initial and main phases of the event and negligible during the late phases of the event. The ionization effect rapidly diminishes in the middle and lower troposphere, where the ion production is comparable to the average
due to GCR or even smaller, i.e. GCRs produce greater quantity of ions than SEPs below some 8 km a.s.l., because the rapidly falling spectra of the latter.

\bigskip 
\begin{acknowledgments}
This work was supported by the Center of Excellence ReSoLVE (project No. 272157) of the Academy of Finland.
\end{acknowledgments}

\bigskip 

\section{References}
\begin{thebibliography}{9}   
\bibitem{Bazilevskaya08}
G. A.Bazilevskaya et al., ``Cosmic ray induced
  ion production in the atmosphere'', Space Science Reviews,
  \textit{137}, 149--173, 2008.

\bibitem{Mironova2015}
I. Mironova et al., ``Energetic particle influence on the Earth's atmosphere'', Space Science Reviews,
  \textit{194}, 1--96, 2015.

\bibitem{Usoskin09}
I.G. Usoskin et al., ``Ionization of the Earth's
  atmosphere by solar and galactic cosmic rays'', Acta Geophysica,
  \textit{57}, 88--101, 2009.

\bibitem{Usoskin11a}
I.G. Usoskin et al., ``Ionization effect of solar particle GLE events in low and middle atmosphere'', Atmospheric Chemistry and Physics,
  \textit{11}, 1979--1988, 2011.

\bibitem{Mishev2013}
A. Mishev et al., ``Ionization effect of
  nuclei with solar and galactic origin in the earth atmosphere during GLE 69
  on 20 January 2005'', Journal of Atmospheric and Solar-Terrestrial
  Physics, \textit{89}, 1--7, 2013. 


\bibitem{Desorgher05}
L. Desorgher et al., ``A GEANT 4 code for computing the interaction of cosmic rays with the
  Earth's atmosphere'', International Journal of Modern Physics A,
  \textit{20}, 6802--6804, 2005.

\bibitem{Usoskin06}
I.G. Usoskin and G.A. Kovaltsov, ``Cosmic ray induced ionization in the
  atmosphere: Full modeling and practical applications'', Journal of
  Geophysical Research, \textit{111}, D21206, 2006.

\bibitem{Velinov2009}
P.I.Y. Velinov et al., ``Model for induced ionization by
  galactic cosmic rays in the Earth atmosphere and ionosphere'', Advances
  in Space Research, \textit{44}, 1002--1007, 2009.

\bibitem{Wissing2009}
J.M. Wissing and M.-B. Kallenrode, ``Atmospheric ionization module
  osnabr{\"u}ck (aimos): A 3-d model to determine atmospheric ionization by
  energetic charged particles from different populations'', Journal of
  Geophysical Research, \textit{114}, A06, 2009.

\bibitem{Mishev13b}
A. Mishev and P.I.Y. Velinov, ``A Maverick GLE 70 in Solar Minimum. Calculations of Enhanced Ionization in the Atmosphere Due to Relativistic Solar Energetic Particles'', Comptes rendus de l'Acad\'emie bulgare des Sciences, \textit{66},  1457--1462, 2013.

\bibitem{Mishev15}
A. Mishev and P.I.Y. Velinov, ``Time evolution of ionization
  effect due to cosmic rays in terrestrial atmosphere during GLE 70'', Journal of Atmospheric and Solar-Terrestrial Physics, \textit{129},  78--86, 2015.

\bibitem{Mishev07}
A. Mishev and P.I.Y. Velinov, ``Atmosphere Ionization Due to
Cosmic Ray Protons Estimated with CORSIKA Code Simulations'', Comptes rendus de l'Acad\'emie bulgare des Sciences, \textit{60}, 225--230, 2007.

\bibitem{Porter76}
H. Porter et al., `` Efficiencies for production of atomic nitrogen and oxygen by relativistic proton impact in air'', Journal of Chemical Physics,
  \textit{65}, 154--167, 1976.

\bibitem{Picone02}
J.M. Picone et al., `` NRLMSISE-00 empirical model of the atmosphere: Statistical comparisons and scientific issues'', Journal of Geophysical Research,
  \textit{107}, 1468, 2002.

\bibitem{Mishev08}
A. Mishev and P.I.Y. Velinov, ``Effects of atmospheric profile variations on yield ionization function Y in the atmosphere'', Comptes rendus de l'Acad\'emie bulgare des Sciences, \textit{61},  639--644, 2008.

\bibitem{Mishev2014}
A. Mishev and P.I.Y. Velinov, ``Influence of hadron and atmospheric models on computation of cosmic ray ionization in the atmosphere-Extension to heavy nuclei'', Journal of Atmospheric and Solar-Terrestrial Physics, \textit{120},  111--120, 2014.

\bibitem{Dryer2001}
M. Dryer et al., ``Prediction in real time of the 2000 July 14 heliospheric shock wave
  and its companions during the 'bastille' epoch'', Solar Physics,
  \textit{204}, 267, 2001.

\bibitem{Klein2001a}
K.-L. Klein et al., ``Coronal electron acceleration and relativistic proton production
  during the 14 July 2000 flare and CME'', Astronomy and Astrophysics,
  \textit{373}, 1073, 2001.

\bibitem{Bom06}
D. J. Bombardieri et al., ``Relativistic proton production during the 2000 July 14 solar event:
  The case for multiple source mechanisms'', Astrophysical Journal,
  \textit{644}, 565, 2006.

\bibitem{Mishev16}
A. Mishev and I. Usoskin, ``Analysis of the Ground-Level Enhancements on 14 July 2000 and 13 December 2006 Using Neutron Monitor Data'', Solar Physics, \textit{292}, 1225--1239, 2016.

\bibitem{Mishev15a}
A. Mishev and P.I.Y. Velinov, ``Ionization rate profiles due to solar and galactic cosmic rays during GLE 59 on Bastille day 14 July 2000'', Comptes rendus de l'Acad\'emie bulgare des Sciences, \textit{68}, 359--366, 2015.

\bibitem{Mishev16a}
A. Mishev and P.I.Y. Velinov, ``Ionization Effect Due to Cosmic Rays during Bastille Day Event – GLE 59 on Short and Mid Time Scales'', Comptes rendus de l'Acad\'emie bulgare des Sciences, \textit{69}, 1479--1484, 2016.

\end{thebibliography}

\end{document}